\documentclass[aps,pre,a4paper,10pt,twocolumn]{revtex4-1}

\usepackage{amsmath}
\usepackage{amsfonts}
\usepackage{amssymb}
\usepackage{amsthm}
\usepackage{mathrsfs}
\usepackage{dsfont}
\usepackage{esint}
\usepackage{graphicx}
\usepackage{subfigure}

\DeclareMathAlphabet{\mathpzc}{OT1}{pzc}{m}{it}

	\newcommand{\AsymEq}{\sim}

	\newcommand{\bb}[1]{\mathbb{#1}}		
	\newcommand{\mr}[1]{\mathrm{#1}}			

	\newcommand{\br}[1]{\left( #1 \right)}
	\newcommand{\brr}[1]{\left[ #1 \right]}
	\newcommand{\brrr}[1]{\left\{ #1 \right\}}
	\newcommand{\of}[1]{\!\br{#1}}
	\newcommand{\off}[1]{\!\brr{#1}}
	\newcommand{\offf}[1]{\!\brrr{#1}}
	\newcommand{\sbr}[1]{( #1 )}
	\newcommand{\sbrr}[1]{[ #1 ]}
	\newcommand{\sbrrr}[1]{\{ #1 \}}
	\newcommand{\sof}[1]{\!\sbr{#1}}
	\newcommand{\soff}[1]{\!\sbrr{#1}}

	\newcommand{\Sum}[2]{\sum\limits_{#1}^{#2}}

	\newcommand{\Int}[3]{\int\limits_{#1}^{#2}\mr{d}#3\,}
	
	\newcommand{\sSum}[2]{\sum_{#1}^{#2}}

	\newcommand{\DA}[1]{\bb{E}\off{#1}}			

	\newcommand{\EA}[1]{\xpc{#1}}

	\newcommand{\xpc}[1]{\left\langle #1 \right\rangle}

	\newcommand{\sEA}[1]{\sxpc{#1}}

	\newcommand{\sxpc}[1]{\langle #1 \rangle}

	\newcommand{\Integers}{\ensuremath{\mathbb{Z}} }

	\newcommand{\landau}[1]{\mathpzc{o}\of{#1}}

		\newcommand{\Min}[2]{\min\of{#1,#2}}

		\newcommand{\sAbs}[1]{\vert #1 \vert}

		\newcommand{\Gma}[1]{\Gamma\of{#1}}
		\newcommand{\sGma}[1]{\Gamma\sof{#1}}

		\newcommand{\PolyLog}[2]{\mathrm{Li}_{#1}\of{#2}}

\usepackage{color}

\begin{document}
	\title{Reaction-diffusion on random spatial networks with scale-free jumping rates via effective medium theory}
	\author{Flavio Iannelli}
	\email{iannelli.flavio@gmail.com}
	\affiliation{Humboldt-Universit\"at zu Berlin, Institut f\"ur Physik, Newtonstra{\ss}e 15, 12481 Berlin}
    \author{Igor M. Sokolov}
    \affiliation{Humboldt-Universit\"at zu Berlin, Institut f\"ur Physik, Newtonstra{\ss}e 15, 12481 Berlin}
    \author{Felix Thiel}
	\email{thiel@posteo.de}
	\affiliation{Bar-Ilan University, Department of Physics, 5290002 Ramat Gan, Israel}
	
\begin{abstract}
		We study epidemic processes using a metapopulation approach on the line featuring random transport rates between arbitrarily distant sites.
		An average transport network is found using a recently developed variant of the effective medium approximation (EMA) that is capable of dealing with these long-range connections.
		Using a Feynman-Kac argument in the effective medium, we derive an estimate on the size of the infected domain, and reproduce the known result of its exponential growth in time.
		We hereby demonstrate the applicability of long-range EMA to dynamical processes on networks more intricate than simple diffusion.
\end{abstract}

\maketitle

\section{Introduction}
\label{sec:Intro}
		Network science has emerged in recent years as a fundamental theoretical framework for the modeling and understanding of large complex systems with many interdependent subunits \cite{barabasi2016network}.
		Virtually any relation or interaction among any set of agents can be represented as a graph.
		Previous works have considered either the structural properties of the network itself \cite{albert2002statistical,boccaletti2006complex} or certain dynamics placed on the network, e.g. coupled oscillators \cite{pikovsky2003synchronization,arenas2008synchronization}, or diffusive transport \cite{noh2004random,masuda2017random,klages2008anomalous,havlin1987diffusion,brockmann2006scaling,brockmann2008anomalous}.  
		The latter is the focus of this paper, in particular we will consider the transport of infectious pathogens.

Understanding the spread of emergent infectious diseases in the geographic space is of fundamental importance in an increasingly connected world. 
In ancient times, the spreading of epidemics, such as the black death, could be understood in terms of diffusive processes \cite{barabasi2016network}.
In those cases the disease is spread by the agents/hosts that can only travel with bounded velocities  between neighboring locations.
This gives rise to a \emph{wave-front} of infected individuals, which travels at a finite speed.
The recent great increase of the connectivity among densely populated areas and the correspondent urbanization, has increased the risk that infectious diseases will spread.  
The complexity of human mobility at all scales, being that urban and inter-urban or world-wide,  is reflected in the possibility for the infection to cross arbitrary distances in close to no time. 
As a consequence, the number of infected sites grows exponentially fast, as opposed to linearly.
Similar phenomena are also discussed in a different biological context, see \cite{hallatschek2014acceleration} and references therein.

		Network theory's success stems not only from its versatility, but even more from the fact that many dynamical processes can be characterized and understood from the underlying connectivity properties of the graph \cite{pastor2001epidemic,boguna2003absence,colizza2007invasion,dorogovtsev2008critical,cohen2000resilience,buldyrev2010catastrophic}.	
		Often, exact knowledge of the network connectivity eludes the theorist, because they are either quickly changing -- as is the case in temporal networks \cite{holme2012temporal} -- or are hard to assess.
		In this case a practical approach is to model our ignorance with a random network.
        
		Many techniques have been developed to deal with dynamical processes on random networks, among them the heterogeneous mean field \cite{pastor2015epidemic} and the annealed adjacency matrix approximation \cite{guerra2010annealed}.
		The main rationale of statistical physics applies: many dynamical details of random networks are determined by a few parameters of the whole {\em ensemble} \cite{albert2002statistical}.
        
		In order to properly understand transport processes it is important to embed the network into the geographic space, i.e. one has to consider spatial networks \cite{barthelemy2011spatial,balister2018topological}.
		The simplest spatial networks are of course lattices.
		In the context of percolation theory, the so-called effective medium approximation has been developed to describe the diffusion on disordered -- i.e. random -- lattices \cite{Choy1999-0,KirkPatrick1973-0}.
		The idea is to replace the initially random transport rates between the nodes with fixed deterministic ones.
		This deterministic average network is called the effective medium, it is characterized by an effective diffusion coefficient.
		EMA is not a blind average of the transport rates, rather it is determined in a self-consistent manner.
		Would a link in the effective medium be replaced by its random original, the transport flux along this particular link would not change on average.
		Hence EMA is particularly suited for systems with independent links.
		In this paper, we employ EMA for infection spreading in the global human traffic network.

		The spatial embedding of the network is especially important in global human traffic, as two topologically adjacent nodes (e.g. airports) may be geographically very far apart.
		Crucially, empirical observations show that human mobility lacks a definite scale  \cite{brockmann2006scaling,gonzalez2008understanding} and features {\em long-range} connections which have been a major limitation for EMA.
		Recently \cite{Thiel2016-1}, we have developed an EMA variant that overcomes this restriction and provides an {\em analytical} technique to deal with random spatial networks -- a model of tremendous complexity.
		The goal of this paper is to demonstrate that and how EMA can be used in reaction-diffusion systems on random spatial networks with long-range connections.
		Contemporary fields where our proposed theory may become relevant are epidemic spreading in the global mobility network \cite{rvachev1985mathematical,colizza2006role,Brockmann2013-0,iannelli2017effective,GomezGardenes2018-0} or dispersal phenomena in biological contexts \cite{hallatschek2014acceleration}.

		For the remaining of the paper we are concerned with epidemic processes in a metapopulation, where the subpopulations are placed on an (ideally) infinite line with random long-range transport rates between them.
        The metapopulation approach has been successfully used to describe spatially embedded subpopulations, such as cities and urban areas, interacting with each other \cite{van2011gleamviz}. 
		Here, we assume diffusive coupling between the subpopulations.
		The individuals travel in the metapopulation and forget about their original subpopulation at each time step. 
		Thus the model is Markovian on the metapopulation level.
		Each individual performs a random walk over the subpopulations with jumping probabilities that are given according to a travel rate matrix.

This gives rise to a network description of the connections between the subpopulations.
		We leverage the Feynman-Kac formula to derive a bound for the diameter of the infected region in a deterministic model.
		Then we proceed to show that this estimate is as well realized in random models, where it can be computed from the effective medium approximation.

		The rest of the paper is organized as follows: We start with section \ref{sec:DetermSR} where we review how to obtain the infection diameter in a deterministic short-range metapopulation model.
		We proceed to review the necessary amendments for deterministic long-range models in section \ref{sec:DetermLR} and finally consider random long-range models in section \ref{sec:RandomLR}.
		This is also where we explain the effective medium approximation.
		Numerical confirmation of our theory is presented in section \ref{sec:num}.
		Discussion and concluding remarks are found in section \ref{sec:disc}.

	\section{Ballistic spreading in deterministic short-range systems}
	\label{sec:DetermSR}
		As an introductory example, we consider the susceptible-infected-susceptible (SIS) metapopulation model on a line.
		Each lattice site in the metapopulation is occupied by a subpopulation, each with the same population size.
		The density of infected individuals at site $x$ and time $t$ is denoted by $\rho_x\sof{t}$, where $x\in\Integers$ and $t\ge 0$.
		The column vector of all $\rho_x$ is denoted without the subscript, i.e. $\rho = \sbr{ \hdots, \rho_{x-1}, \rho_x, \rho_{x+1}, \hdots }^T$.
		Individuals travel to an adjacent subpopulation with a constant transport rate $W$, i.e. $W\rho_x$ infected will travel from $x$ to $x+1$ and to $x-1$, per unit time.
		We assume that these travel rates are symmetric, so that the total number of individuals in each subpopulation does not change in time.
		The change in density of infected individuals can be expressed in terms of the following continuous-time reaction-diffusion equations:
		\begin{equation}
			\dot{\rho}_x\sof{t}
			=
			\sbrr{ \Omega \rho\sof{t} }_x 
			+ \rho_x\sof{t} f[\rho_x\sof{t}]
			.
		\label{eq:SIS}
		\end{equation}
		
		The matrix $\Omega$ is the \textit{transport operator} that describes jumps between adjacent lattice sites \cite{rvachev1985mathematical}.
		In the current example it is equal to the graph Laplacian of the line: 
		\begin{equation}
			\Omega_{x,y}
			:=
			W\delta_{x,y-1}
			+ W\delta_{x,y+1}
			- 2 W \delta_{x,y}
			.
		\label {eq:DefSRTransOp}
		\end{equation}
        
        Initially we infect a fraction $c_0$ of the site on the origin, i.e. $\rho_x\sof{t=0} = c_0 \delta_{x,0}$.
		Locally in each subpopulation, the infection dynamics take place.
		It is described by the reaction term
		\begin{equation}
			\rho_x f[\rho_x\sof{t}]
			:=
			\beta \rho_x \sbr{ 1 - \rho_x }
			- \mu \rho_x 
			.
		\label{eq:DefReactionTerm}
		\end{equation}
		The second term describes the recovery of infected individuals with rate $\mu$, $I \overset{\mu}{\to} S$, when the infected become susceptible again.
		As the total number of individuals per subpopulation is conserved, the density of susceptible individuals is given by $1-\rho_x\sof{t}$.
		This reveals the first term in Eq.~\eqref{eq:DefReactionTerm} as the infection of a susceptible individual with rate $\beta$, $S + I \overset{\beta}{\to} 2 I$.
		Note that the local reaction rate is bounded from above by 
		\begin{equation}
			f[\rho_x\sof{t}]
			\le 
			\beta - \mu
			.
		\label{eq:ReactionTermBound}
		\end{equation}
		The ratio $\mathcal{R}_0 = \beta/\mu$ is the basic reproductive number and denotes the average number of secondary infections caused by a primary case in a fully susceptible population.
		The infection can be sustained locally in the long-time limit only when $\mathcal{R}_0 > 1$, or equivalently when $\beta > \mu$, which we will assume throughout the text.
		In this case a single infected agent in an otherwise susceptible population will lead to a steady state with a non-zero infection density given by $(\beta-\mu)/\beta$ \cite{barrat2008dynamical}.
		A sketch is given in Fig. 1 where we show the interplay between reaction and diffusion in the metapopulation model.
		\begin{figure*}
		\includegraphics[width=1.99\columnwidth]{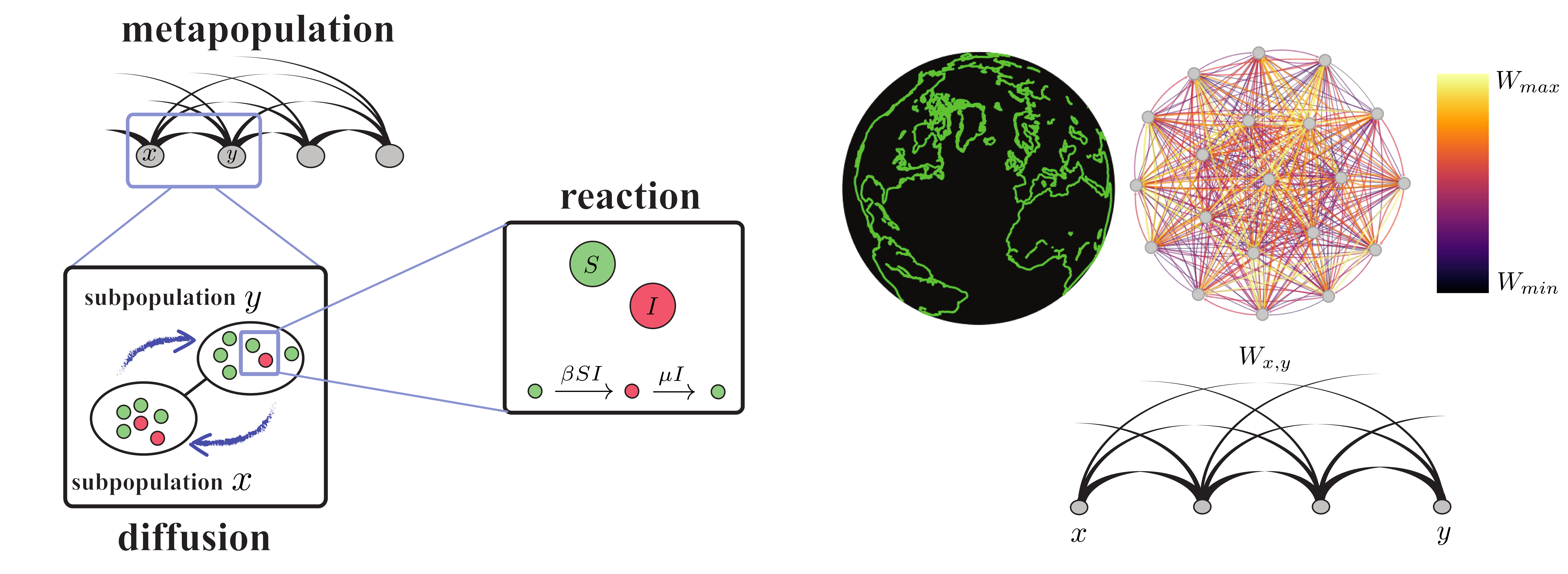}
		\caption{
			(Color online) Left panel: scheme of the one-dimensional metapopulation model. The dynamics is is regulated by two different time scales, the one of diffusion, corresponding to the supopulation layer, and the reaction, governed by the SIS infection dynamics at the  individual layer. Right panel: illustration of a sample metapopulation network consisting of $N=20$ subpopulations with symmetric transition rates $W_{x,y}$. The graph is constructed from a one-dimensional ring topology with all connections permitted, thus allowing the mapping with a two-dimensional space such as the earth geographical space. The edge color and size scales accordingly with the values of each transition rate.
			\label{fig:metapop}
		}
		\end{figure*}

		One way to investigate the spreading process is to apply the Feynman-Kac representation  to solve Eq.~\eqref{eq:SIS}.
		This approach was used e.g. in Ref.~\cite{Mancinelli2003-0}.
		Here, one considers a random walk $X\sof{t}$, with $X\sof{t=0} = 0$ whose pdf $P_x\sof{t}$ is determined by Eq.~\eqref{eq:SIS} without the reaction term:
		\begin{equation}
			\dot{P}_x\sof{t}
			=
			\sbrr{\Omega P\sof{t}}_x
			.
		\label{eq:Diff}
		\end{equation}
		Then the solution of Eq.~\eqref{eq:SIS} is given by the following implicit equation:
		\begin{equation}
			\rho_x\sof{t}
			=
			\EA{
				\rho_{x+X\sof{t}}\sof{0}
				\exp\brr{ \Int{0}{t}{t'} f[\rho_{x+X(t')}\of{t'}] }
			}
			,
		\label{eq:FeynmanKac}
		\end{equation}
		where the average is taken with respect to the random-walk realizations $\{X\sof{t}\}$.
		The Feynman-Kac equation relates the characteristic function of certain integrals of random processes with a partial differential equation and vice versa.
		The rationale is the same like in the construction of path integrals in quantum mechanics.
		Although it is hard to solve Eq.~\eqref{eq:FeynmanKac} explicitly, it is easy to find an upper bound.
		This is done by replacing the exponent via Eq.~\eqref{eq:ReactionTermBound}, by plugging in the initial condition $\rho_x\sof{0} = c_0\delta_{x,0}$ and by recognizing $\sEA{\delta_{X\sof{t},x}} = P_x\sof{t}$:
		\begin{equation}
			\rho_x\sof{t} 
			\le 
			c_0 e^{\sbr{\beta-\mu}t}
			P_x\sof{t}
			\AsymEq 
			\frac{
				c_0 
			}{\sqrt{4 \pi W t}}
			e^{\sbr{\beta-\mu}t - \frac{x^2}{4Wt}}
			.
		\label{eq:FundIneqSR}
		\end{equation}
		For the second part of the expression we used the fact that $P\sof{t}$ approaches a Gaussian for large times.
		An alternative way to derive the inequality is to linearize Eq.~\eqref{eq:SIS}.

		The inquality~\eqref{eq:FundIneqSR} is useful, when one considers the $\bar{c}$-level set of the infected fraction, i.e. all sites $x$, such that $\rho_x\sof{t} \ge \bar{c}$.
		Chaining the inequalities, one obtains 
		\begin{equation}
			\bar{c} 
			\le
			\frac{
				c_0 
			}{\sqrt{4 \pi W t}}
			e^{\sbr{\beta-\mu}t - \frac{x^2}{4Wt}}
			,
		\label{eq:}
		\end{equation}
		which can be solved for $x$ and yields a (time-dependent) radius of the infected region:
		\begin{equation}
			x
			\le 
			2 \sqrt{\sbr{\beta-\mu} W} t
			+ \landau{t}
			.
		\label{eq:}
		\end{equation}
		The {\em diameter} of the infected region is twice of the above radius and is asymptotically bounded:
		\begin{equation}
			D\sof{t} 
			\le 
			4\sqrt{ \sbr{\beta-\mu} W} t.
		\label{eq:ball}
		\end{equation}
		This shows that for large times the infection spreads no faster than  {\em ballistically} \cite{fisher1937wave,tikhomirov1991study}, with a velocity that grows monotonically with the transport rate $W$, which is indeed the case \cite{Belik2011-0}.
		For a diffusion-limited infection this means that there is an upper bound for the front propagation speed.
		This is a consequence of the Gaussianity of $P_x\sof{t}$ (see the discussion in \cite{Mancinelli2003-0}) which in turn is related to the lack of long-range connections.

		With fixed reaction dynamics, the above reasoning can be extended in two directions: 
		(i) the introduction of transport beyond the nearest-neighbour population, and/or
		(ii) make the transport rates a random quantity.
		This will be done in the next two sections.

	\section{Exponential spreading in deterministic long-range systems}
	\label{sec:DetermLR}
		To model the fast multi-scale human mobility, one might consider the introduction of more than nearest neighbor connections in the transport operator.
		Instead of Eq.~\eqref{eq:DefSRTransOp} one might consider:
		\begin{equation}
			\Omega_{x,y}
			=
			\sbr{ 1 - \delta_{x,y} } W_{x,y} 
			+
			\delta_{x,y} \Sum{z \ne x}{} W_{x,z}
			,
		\label{eq:DefLRTransOp}
		\end{equation}
		where the transition rates $W_{x,y}$ are symmetric, i.e. $W_{x,y} = W_{y,x}$,  and decay with the distance, so that the sum in the diagonal terms of $\Omega$ converges.

		Consider first the example, when the transport rates decay like a power law with distance:
		\begin{equation}
			W_{x,y}
			= 
			\frac{K}{\sAbs{x-y}^{1+\alpha}}
			,
		\label{eq:LevyRates}
		\end{equation}
		with $\alpha \in (0,2)$ and where $K$ plays the role of an anomalous diffusion constant.

		All reasoning from section~\ref{sec:DetermSR} can be repeated up to Eq.~\eqref{eq:FundIneqSR}.
		However, the random walk generated by the transport operator of Eq.~\eqref{eq:DefLRTransOp} with rates given by Eq.~\eqref{eq:LevyRates}, is very different from before.
		Due to the possibility of long-range jumps that lack a finite variance, it will not converge to a Brownian motion, but instead to an $\alpha$-stable distribution which is characterized by power-law tails instead of a Gaussian decay \cite{Feller1971-0}:
		\begin{equation}
			P_x\sof{t}
			\AsymEq
			\frac{\alpha K t}{\sAbs{x}^{1+\alpha}}
			.
		\label{eq:EMAAsym}
		\end{equation}
		These scale-free random walks are known as L\'evy flights, and $\alpha \in (0,2)$ is the L\'evy exponent. A derivation of the previous equation is reproduced in Appendix \ref{app:Prop}.
		Using this power law in Eq.~\eqref{eq:FundIneqSR} and solving for $\sAbs{x}$ allows us to estimate the diameter of the infected region:
		\begin{equation}
			D\sof{t}
			\le 
			2
			\br{ \alpha K \frac{c_0}{\bar{c}} t}^{\frac{1}{1+\alpha}}
			e^{\frac{\beta-\mu}{1+\alpha} t}
			,
		\label{eq:EMADiameter}
		\end{equation}
		which, contrary to the ballistic growth ~\eqref{eq:ball} found for bounded jumps, grows exponentially fast.
        
		One might argue that the assumed power law decay in the transition rates is rather specific and far off the measured travel rates.
		In order to overcome the this problem, we model our ignorance about the actual travel rates with chance.

	\section{Exponential spreading in random long-range systems}
	\label{sec:RandomLR}
		We now consider the reaction diffusion equation~\eqref{eq:SIS} with a transport operator \eqref{eq:DefLRTransOp} that features {\em random} rates $W_{x,y}$.
		It is assumed that the rates are random variables independently placed on each link $\sbr{x,y}$.
		They are symmetric $W_{x,y} = W_{y,x}$ and decay with the distance between the nodes $\sAbs{x-y}$ such that the diagonal terms in Eq.~\eqref{eq:DefLRTransOp} are well defined.
		Hence, there is a family $\sbrrr{p_{\sbr{x,y}}\sof{w}}$ of probability density functions, that describe the distribution of $W_{x,y}$ and that in total describes the ensemble of random networks.

		The Feynman-Kac equation could still be used, but it would involve the random walk in a random network generated by the random operator $\Omega$.
		Since this is a rather hopeless venture, we will first employ EMA to compute an average transport operator $\widetilde{\Omega}$.
		For details we refer to Ref.~\cite{Thiel2016-1}.
		We call the average network described by $\widetilde{\Omega}$ the effective medium.
		The deterministic rates $\widetilde{W}_{x,y}$ of the EMA operator have to be chosen such, that 
		(i) any link that is present in some network of the ensemble will be present in the effective medium, albeit with possibly different strength; 
		and such that (ii) the distance scaling in $p_{\sbr{x,y}}\sof{w}$ is preserved.
		These conditions are necessary for the effective medium to be well defined.
		The transport rates are determined by the following set of self-consistency equations:
		\begin{equation}
			0
			=
			\DA{
				\frac{
					\widetilde{R}_{x,y} \br{W_{x,y} - \widetilde{W}_{x,y} }
				}{ 
					1 + \widetilde{R}_{x,y} \br{ W_{x,y} - \widetilde{W}_{x,y} } 
				}
			}
			.
		\label{eq:EMA}
		\end{equation}
		Here, the average is taken over the distribution of one fixed transport rate $W_{x,y}$.
		$\widetilde{R}_{x,y}$ is the so-called resistance distance \cite{Bapat2014-0} computed from the (pseudo-)inverse of $\widetilde{\Omega}$:
		\begin{equation}
			\widetilde{R}_{x,y}
			:=
			\widetilde{\Omega}^{-1}_{x,y}
			+ \widetilde{\Omega}^{-1}_{y,x}
			- \widetilde{\Omega}^{-1}_{x,x}
			- \widetilde{\Omega}^{-1}_{y,y}
			.
		\label{eq:DefResDist}
		\end{equation}
		The expression in Eq.~\eqref{eq:EMA} describes the average change in the stationary transport flux upon replacement of the effective medium link along $\sbr{x,y}$ with its random original $W_{x,y}$.
		EMA requires this change to vanish on average.
		For this reason it is very successful in reproducing the diffusive properties of the random network ensemble.
		It is important to note that the actual choice of the effective medium graph is mostly arbitrary, as long as the two conditions given above are respected.

		Eq.~\eqref{eq:EMA} constitutes a set of equations for each class of links that share the same distribution.
		It simplifies considerably, if one assumes scaling behavior between distance and rates.
		We will focus here on the simplest case, when the transport rates are given by some i.i.d. random number divided by a power of the distance
		\begin{equation}
			W_{x,y}
			=
			\frac{Z_{x,y}}{\sAbs{x-y}^{1+\alpha}}
			,
		\label{eq:RateAss}
		\end{equation}
		where $\alpha \in (0,2)$ and $Z_{x,y}$ is a family of i.i.d. random variables.
		In Ref.~\cite{Thiel2016-1} it was shown that the actual distribution of the rescaled transition rates $Z_{x,y}$ does not influence the qualitative behavior of the effective medium as the effective medium transition rates are given by:
		\begin{equation}
			\widetilde{W}_{x,y}
			=
			\frac{\DA{Z_{x,y}}}{\sAbs{x-y}^{1+\alpha}}
			.
		\label{eq:EMARates}
		\end{equation}
		As long as the mean transition rate is finite, the effective medium is exactly the deterministic long-range system of sec.~\ref{sec:DetermLR} with $K = \DA{Z_{x,y}}$.
		Recently, it was proven under certain regularity conditions that this is the correct self-averaging limit of the random walk in the random network \cite{Chen2018-0}.
		We can draw the same conclusions for the random model as we did for the deterministic one, namely that the diameter of the infected regions grows exponentially, just like in Eq.~\eqref{eq:EMADiameter}.

		Although, this behavior is known in the literature \cite{Mancinelli2003-0,del2003front,brockmann2007front}, EMA opens a new way to {\em analytically} compute the speed of the infection spreading or even other quantities of desire.
		Importantly,  the method presented here is not limited to the simple topology and the simple choice of transport rates that we used in Eq.~\eqref{eq:RateAss}.
		In our example, the effective medium transport rates are simple averages of the original rates and the EMA result becomes equal to the annealed adjacency matrix approximation of Ref.~\cite{guerra2010annealed}.
		This is however a consequence of the high connectivity and the power-law in Eq.~\eqref{eq:RateAss} and doeas not have to hold in general.
		For more general topologies or other scaling relations, a different effective medium has to be chosen.
		This is already seen in the traditional EMA examples, e.g. a random short-range model (the so-called random barrier model, see e.g. \cite{Bouchaud1990-0}), where only next-neighbor transport is allowed.
		The equation system \eqref{eq:EMA} reduces to a single equation for the effective medium diffusivity $K$:
		\begin{equation}
			0 
			= 
			\DA{
				\frac{
					K - W_{x,x+1}
				}{
					W_{x,x+1} + \sbr{d-1} K
				}
			}
			.
		\label{eq:}
		\end{equation}
		For a barrier model in one dimension $d=1$, one finds $K = \DA{1/W_{x,x+1}}^{-1}$.
		The effective medium diffusivity is given by the reciproke of the harmonic mean, instead of by the arithmetic mean of the transport rates.

		Since EMA reproduces the diffusive behaviour of random systems pretty well \cite{Thiel2016-1}, it is a good candidate to produce a disorder-averaged random walk that can be used in Eq.~\eqref{eq:FeynmanKac}.
		Using EMA, one can make predictions about the reaction-diffusion system with a {\em random} transport operator, as we demonstrate numerically in the next section.

\section{Numerical results}        
\label{sec:num}
	To validate our theory, we consider a ring of subpopulations with
	transport rates defined by Eq.~\eqref{eq:RateAss}.
	As mentioned above, the actual distribution of $Z_{x,y}$ does not matter, hence we sampled them uniformly from the interval $[0,1]$.
	Therefore $K = \DA{Z} = 0.5$ in our simulations.
	It is important to note that the metric of a ring in one dimension is used
	\begin{equation}
	\sAbs{x-y} = \Min{\sAbs{x-y}}{N-\sAbs{x-y}}.
	\end{equation}
	This determines the upper triangle of $\Omega$; the lower triangle is given by the symmetry condition of $\Omega$.
	Its diagonal elements are the negative sum of all other elements in the respective column.
	With this random transport operator, Eq.~\eqref{eq:SIS} is integrated using a fifth order Runge-Kutta method.
	For each realization of $\Omega$ we obtain a collection of $\rho_x\of{t|\Omega}$.
	Then we computed the average $\rho_x\of{t} = \DA{\rho_x\of{t|\Omega}}$ over 50 realizations of $\Omega$.
	Given the infection threshold $\bar{c} = 0.1$ we compute the infection diameter via:
	\begin{equation}
		D\sof{t}
		:=
		\mathrm{diam}\offf{x| \rho_x\sof{t} \ge \bar{c}}
		.
	\label{eq:}
	\end{equation}

\begin{figure}
		{\includegraphics[width=0.78\columnwidth]{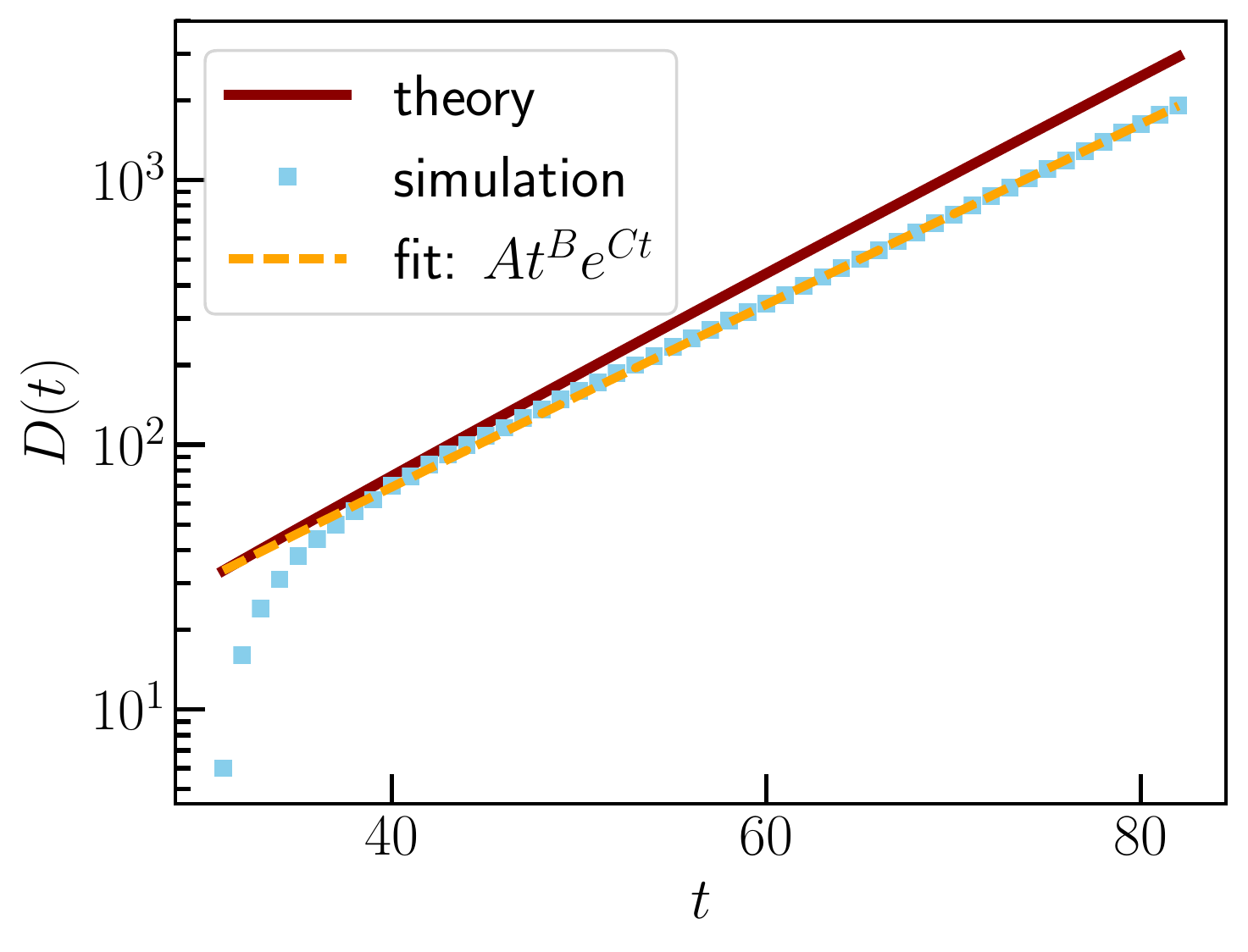}}
		\caption{
			(Color online) Diameter of the infected population obtained from numerical integration of the reaction-diffusion equations and the EMA prediction given by the upper bound of Eq.~\eqref{eq:EMADiameter}. The epidemic is generated by a subpopulation SI reaction with $\beta=0.2$ in $N=4000$ subpopulations with L\'{e}vy exponent $\alpha=1.5$. 
			\label{fig:diam}
		}
	\end{figure}

	Initially,  we consider a simple susceptible-infected (SI) reaction scheme ($\mu=0$) with $\beta=0.2$ and $\alpha=1.5$ in $N=4000$ supopulations  with initial concentration of infected at the origin $c_0 = 10^{-2}$.
	A comparison of $D\sof{t}$ with the upper bound in Eq.~\eqref{eq:EMADiameter} is given in Fig.~\ref{fig:diam}.
	The numerical data respects the bound nicely.


	Since the numerical diameter $D\of{t}$ shows a nice exponential growth pattern like in Eq.~\eqref{eq:EMADiameter}, we can extract some of the parameters from the exponential fit
	\begin{equation}
		D\of{t} = A t^B e^{Ct}.
	\label{eq:Fit}
	\end{equation}
	Comparison with Eq.~\eqref{eq:EMADiameter} would give measured values for $\alpha$, $\beta - \mu$ and the diffusivity $K = \DA{Z_{x,y}}$.
	This may however be a hard task, because the non-linear term $t^B$ is not easy to detect in the exponential 
	fit.
	The diameter's growth rate
	\begin{equation}
		C
		:= 
		\frac{\beta-\mu}{1+\alpha}
	\label{eq:DefGrowthRate}
	\end{equation}
	on the other hand is easy to obtain, as it can also be measured from the slope of the tail of $\ln D\of{t}$.
	In our simulations of the SI metapopulation model we obtain $C = 0.076$, which gives $\alpha_{fit} = 1.622$. 
	This is a reasonably  close value to the L\'{e}vy exponent $\alpha = 1.5$ used for generating the graph realizations in the first place.
	$D\sof{t}$ is only presented before the saturation sets in, and before the whole ring is infected.
    
\begin{figure}
		{\includegraphics[width=0.78\columnwidth]{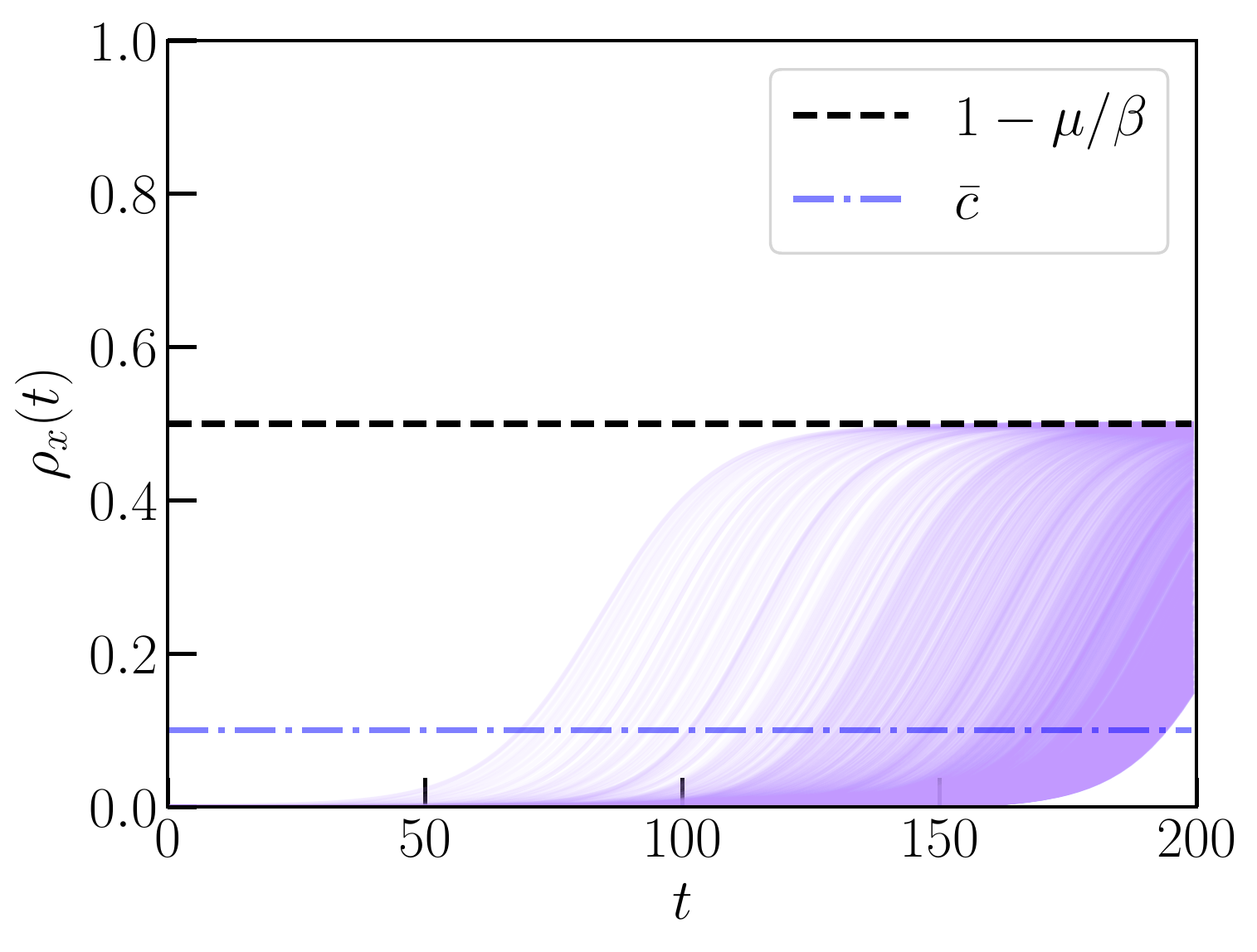}}
		\caption{
			(Color online) Prevalence curves (violet) for the SIS reaction  with $\beta=0.2$ and $\mu=0.1$ of the $N=8000$ fully connected subpopulations with rates' L\'{e}vy exponent $\alpha=1.5$. The asymptotic value of the SIS steady state, the disease prevalence $\rho_x(\infty) = (\beta-\mu)/\beta$, is marked by the black dashed line while the concentration threshold $\bar{c}$ that defines the infection outbreak in each subpopulation is marked by the dash-dot blue line.
			\label{fig:SIS_prevalence}
		}
	\end{figure}

	We now consider the SIS model in $N=8000$ subpopulations with $\beta=0.2$ and $\mu=0.1$, which gives a basic reproductive number of $\mathcal{R}_0 = 2$. 	
	The correct time frame to assess $D\sof{t}$ is visible in a prevalence plot, see Fig.~\ref{fig:SIS_prevalence}.  
	In this figure the curves $\rho_x\sof{t}$ for each $x$ are plotted against time; the stationary value $\rho_x\sof{\infty} = 1 - \mu/\beta$ as well as the time when $\rho_x\sof{t} > \bar{c}$ can be read from such a plot.
	For $N=8000$ subpopulations the time gap between the outbreaks of the first and last subpopulation infected is $124$ time steps, and the absolute global infection time is $193$ time steps. 
	As expected, the results are similar to the SI case.
	For the estimation of the L\'{e}vy exponent at this reproductive number we find $C = 0.041$, which results in $\alpha_{fit} = 1.454$, i.e. in only $3\%$ error of the theoretical value.

\begin{figure}[]
		\subfigure{\includegraphics[width=0.755\columnwidth]{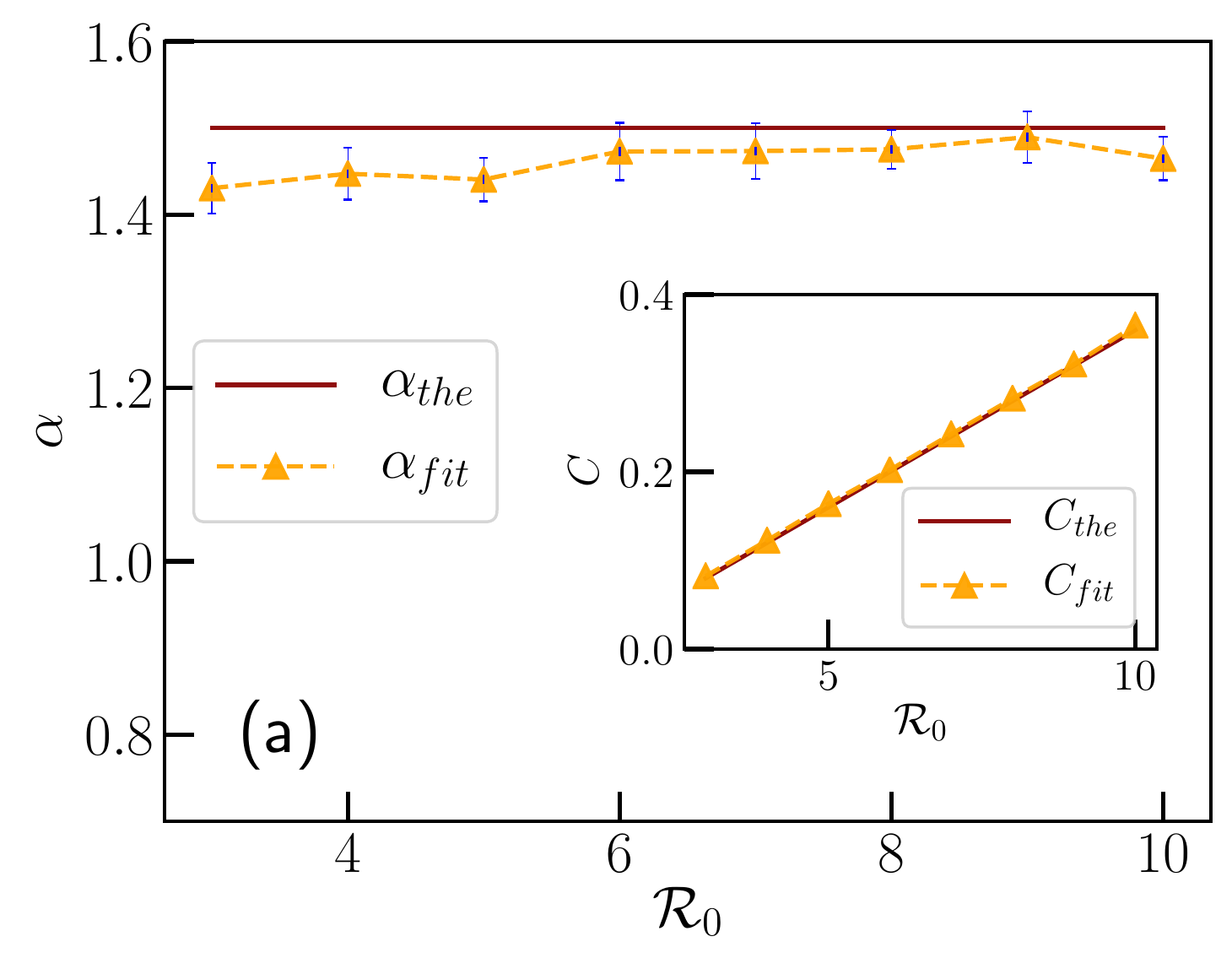}}
		\subfigure{\includegraphics[width=0.78\columnwidth]{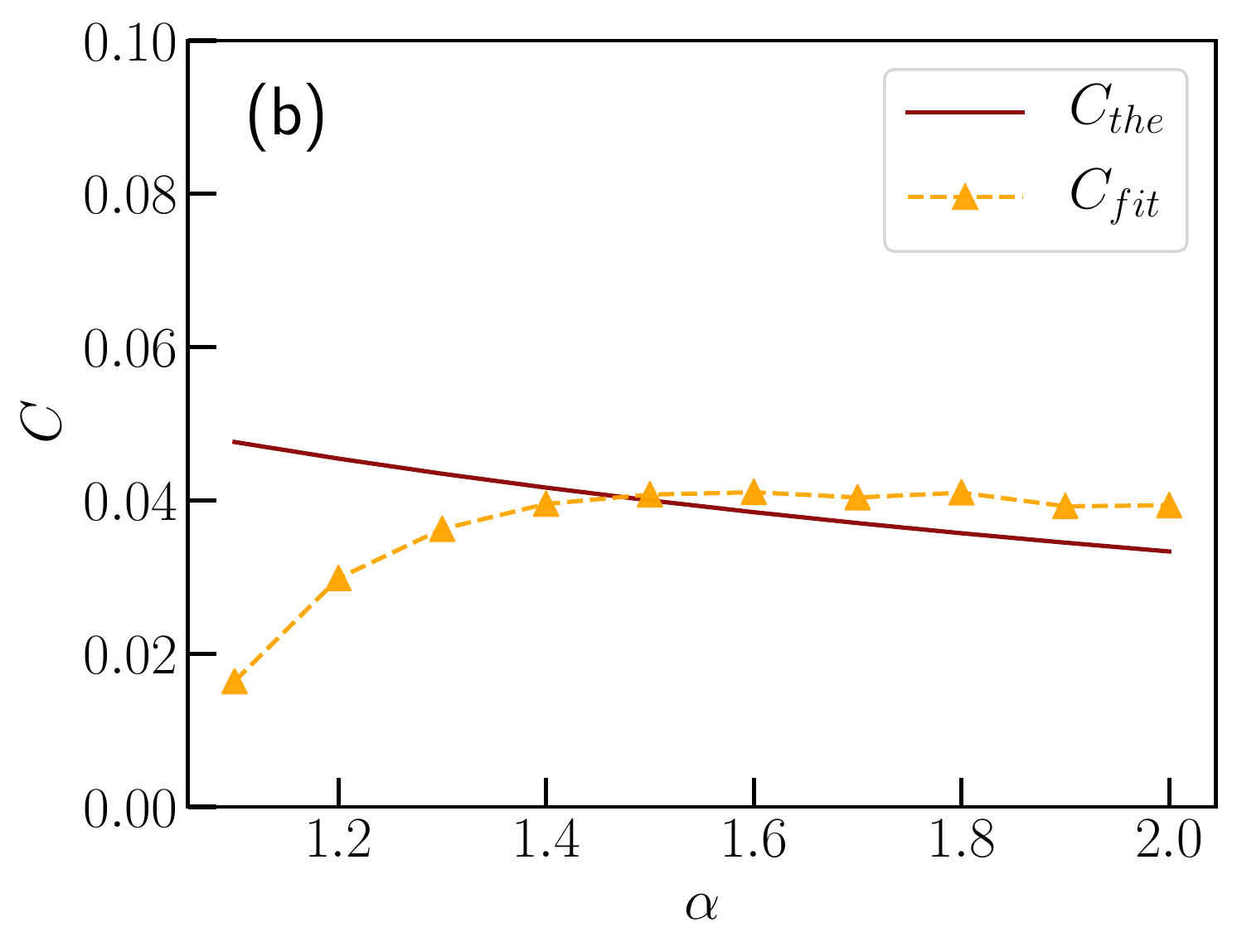}}
		\caption{
         (Color online) The extrapolated value of the L\'evy exponent $\alpha$  with the corresponding error (blue bars) evaluated from the error propagation of the numerical fit error in $C$, shown in the inset,  as a function of the basic reproductive number $\mathcal{R}_0=\beta/\mu$ at fixed $\alpha=1.5$ (a). 
			Theoretical growth rate $C_{the}$ and the respective simulation fit value $C_{fit}$ for the SIS reaction with  $\mu=0.1$ in $N=8000$ subpopulations as a function of the L\'evy exponent $\alpha\in(1,2]$  at fixed $\beta=0.2$ (b).
			\label{fig:Calpha}
		}
	\end{figure}

	Varying the reproductive number and measuring the growth rate $C$ or the L\'{e}vy exponent $\alpha$, respectively, leads to good coincidence between theory and numerics, see Fig.~\ref{fig:Calpha} (a).
	When the theoretical L\'{e}vy exponent $\alpha$ is varied and the growth rate is measured, the agreement appears much worse, see Fig.~\ref{fig:Calpha} (b).

	\begin{figure}[]
		\includegraphics[width=0.94\columnwidth]{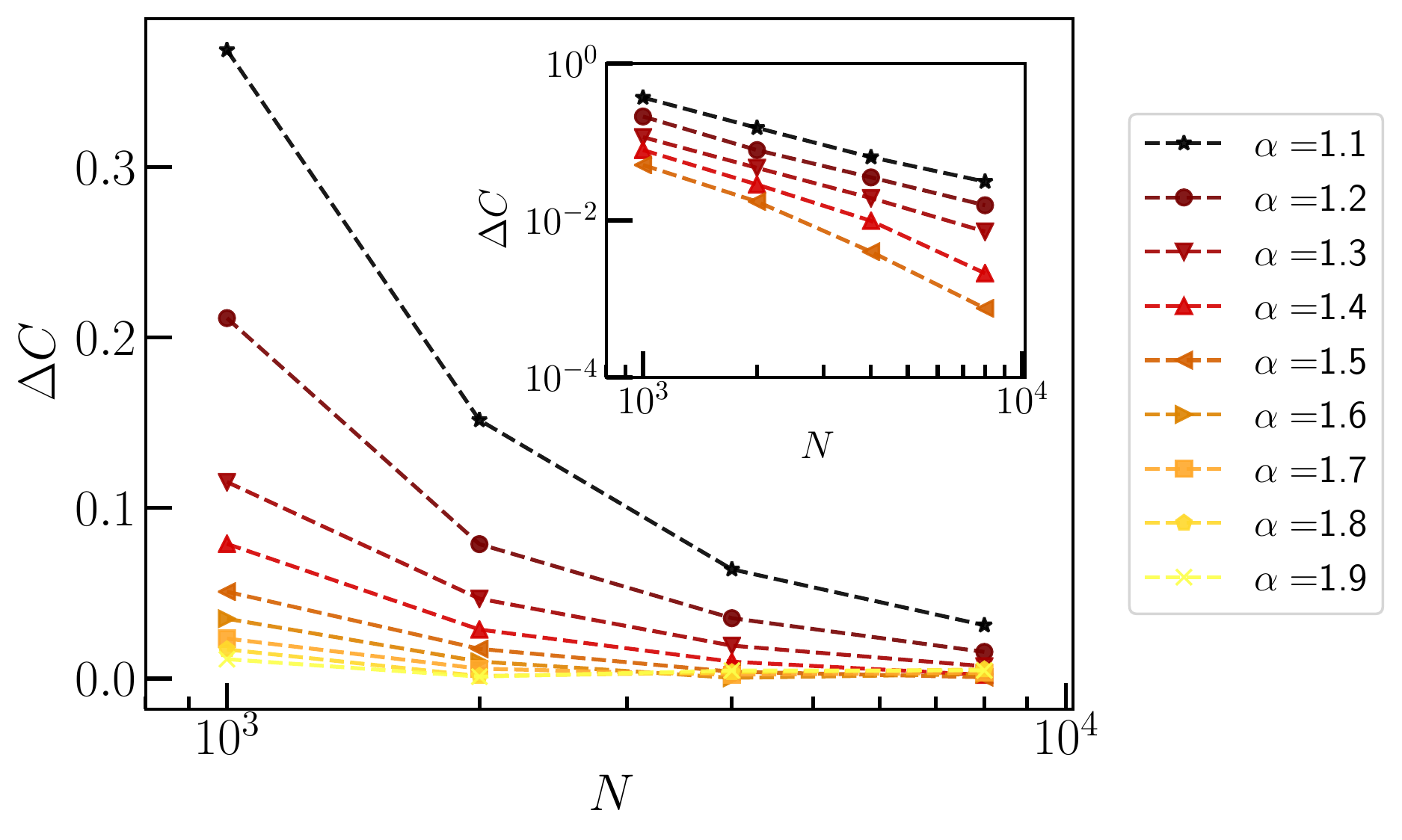}
		\caption{
               (Color online) Absolute value of the difference $\Delta C := \sAbs{C_{the}-C_{fit}}$ between the theoretical $C_{the}=(\beta-\mu)/(1+\alpha)$ and the simulation fit value $C_{fit}$ for the SIS reaction with $\beta=0.2$ and $\mu=0.1$ as a function of the subpopulations number $N$. Different lines are for different L\'{e}vy exponents from dark to light in the range $\alpha\in(1,2)$. Inset: close-up in doubly logarithmic scale for  $\alpha\in(1,1.5)$.
For larger values of $\alpha$, the error fluctuates around $0.005$ which is the numerically attainable accuracy.
			\label{fig:c-c}
		}
	\end{figure}
    
	This mismatch is easily explained as pure finite size effects as we show in Fig.~\ref{fig:c-c}.
	There, we plotted the difference $\sAbs{C_{the} - C_{fit}}$ between the measured growth rate $C_{fit}$ and its theoretical prediction from Eq.~\eqref{eq:DefGrowthRate} in a double logarithmic fashion against the system size $N$.
	The figure shows that the error decays at least like a power law and will vanish in the thermodynamic limit $N\to\infty$.
	Due to the extreme long-range connections, $\rho_x\sof{t}$ saturates very quickly.
	This leads to very short time frame in which $D\sof{t}$ grows exponentially that makes a correct estimation of $C$ difficult.
	The effect becomes worse as $\alpha$ decreases, which also explains the slightly worse agreement for small $\alpha$ in Fig.~\ref{fig:Calpha} (a).
	For this reason we concentrated our numerical studies to the range $\alpha \in (1,2)$.

\begin{figure*}[]
		\includegraphics[width=1.98\columnwidth]{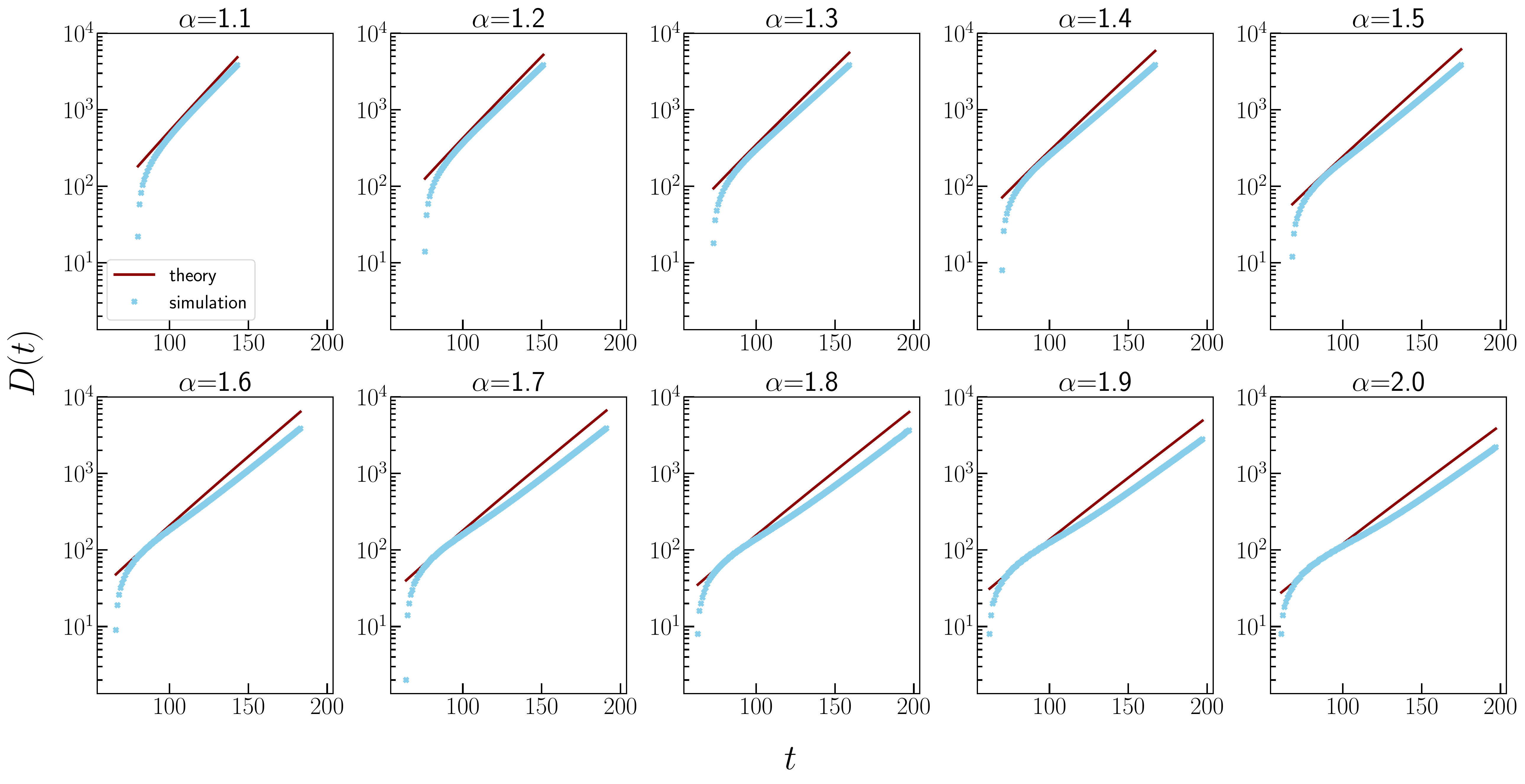}
		\caption{
		(Color online) Diameter of the infected population obtained from the simulation (light-blue scatter) of the SIS  reaction  in $N=8000$ subpopulations with $\beta=0.2$ and $\mu=0.1$. The theoretical prediction (dark-red line) is given by EMA  for various  L\'{e}vy exponent $\alpha\in(1,2]$.
		\label{fig:alpharange}
		}
	\end{figure*}

	We found in Fig.~\ref{fig:Calpha} (b) that $C_{fit} > C_{the}$ in that data range, which should not be possible as Eq.~\eqref{eq:EMADiameter} represents an upper bound.
	An overview of the agreement with the theoretical bound for the probed range in $\alpha$ is shown in Fig.~\ref{fig:alpharange}.
	We find that for some of these large values of $\alpha$ the numerical data overestimate the EMA bound.
	This, however, only happens in an intermediate time regime and not in the long time limit, in which we derived Eq.~\eqref{eq:EMADiameter}.
	In fact, we find that the upper bound is respected in the long-time limit for all values of $\alpha$.
	The predictions given by EMA are rigorously valid in the thermodynamic limit $N\to\infty$, when the infection propagates indefinitely and saturation is never reached.

\section{Discussion}
\label{sec:disc}

The goal of this paper was to present a new analytical tool for reaction-diffusion problems in random long-range networks.
We wanted to advocate the use of effective medium theory that provides a deterministic representative for an originally random network.
Together with the standard Feynman-Kac argument we provide an upper bound for the infection spread in a simple SIS model that is well respected in the long time limit of our numerical simulations.
We also demonstrated that certain parameters, like the L\'evy exponent $\alpha$, can be extracted from data, thus verifying that a made assumption on the random network's ensemble is correct.
This way we demonstrated that EMA is still relevant even beyond the short-range connection paradigm.

With the human travel network in mind, we presented a simple metapopulation model with random long-range connections.
We reproduced the exponential growth of the infection diameter, that is known in the literature \cite{Mancinelli2003-0,del2003front,brockmann2007front}.
Our EMA prediction of the growth rate depends on both the infection and recovery rates $\beta$ and $\mu$ as well as on the topology encoded in the L\'{e}vy exponent $\alpha$ of the statistical decay of the link strength, see Eq.~\eqref{eq:RateAss}.
Other characteristics of the transition rates (like their mean) only play a minor role in the dynamics.
Notice, that long-range links with a ``weak'' power law -- i.e. $\alpha > 2$ in Eq.~\eqref{eq:RateAss} -- would eventually lead to a ballistic growth of the infection front.
These results are also discussed in Ref.~\cite{hallatschek2014acceleration}.


The main restrictions of EMA are currently the necessity of independent and symmetric transport rates $W_{x,y} = W_{y,x}$.
Future modifications of EMA are necessary to deal with asymmetric rates, and can thus take variable subpopulation sizes into account.
Furthermore, when it is possible to deal with {\em correlated} links, more realistic models than a simple grid of the subpopulation's locations can be included.

In its current form, EMA could already be used to tackle more involved models than the one considered here.
E.g. an extension of our argument to $d$ dimensions is possible without major change and would only lead to a different growth rate of $C = (\beta-\mu)/(d+\alpha)$.
Internal dynamics on the nodes (like commuting agents) could be considered by replacing the subpopulations by small networks themselves.
The EMA method is not restricted to the simple model considered here.
In particular, one can overcome the strong finite size effects, that we encountered in our work by considering a finite size effective medium instead of an infinite one, as we did here for simplicity.

EMA is known to nicely reproduce the transport behavior of a random system {\em provided} it is far away from the percolation threshold.
The networks we treated here are very well connected due to the presence of the long-range links.
Therefore they are generically far from percolation threshold, which partly explains the success of our approach. Note that the case discussed in Ref.~\cite{PhysRevLett.79.857} violates both assumptions of absence of correlations and deviation from percolation transition and leads to a very different behavior termed paradoxical diffusion.

Future work and applications of EMA to reaction-diffusion systems include the generalization to arbitrary heterogeneous connectivity networks, such as real-world networks of human mobility.
We believe that EMA will develop to a great practical tool for the analysis of dynamics on networks.

\acknowledgements
The authors are indebted to A. Vulpiani for insightful  discussions.

This work has partially been founded by the DFG / FAPESP, within the scope of the IRTG 1740 / TRP 2015/50122-0.

\appendix
\section{Asymptotic behavior of $\widetilde{P}_x$}
\label{app:Prop}

In this appendix we show how to obtain Eq.~\eqref{eq:EMAAsym} from Eq.~\eqref{eq:LevyRates}.
The computation follows closely Ref.~\cite{Thiel2016-1}.
We start by plugging the transport rates into Eq.~\eqref{eq:Diff} and using their symmetry to reorder the summation:
\begin{equation}
\dot{\widetilde{P}}_{x} = K \Sum{\xi=1}{\infty} 
\frac{\widetilde{P}_{x+\xi}\of{t} + \widetilde{P}_{x-\xi}\of{t} - 2\widetilde{P}_x\of{t}}{\xi^{1+\alpha}}
.
\end{equation}
The equation is solved using Fourier transform, i.e. we multiply $e^{ikx}$ on both sides and sum overall $x$.
Defining $\widetilde{P}\of{k;t} = \sSum{x\in\Integers}{} e^{i k x} \widetilde{P}_x\of{t}$, we obtain:
\begin{align}
	\dot{\widetilde{P}}\sof{k;t}
    = &  \nonumber
    K
    \Sum{x\in\Integers}{}
    \Sum{\xi=1}{\infty}
    e^{ikx}
  	\frac{
    	\widetilde{P}_{x+\xi}\of{t} 
        + \widetilde{P}_{x-\xi}\of{t} 
        - 2\widetilde{P}_x\of{t}
    }{\xi^{1+\alpha}}
	\\ = & \nonumber
    K
    \Sum{\xi=1}{\infty}
    \frac{1}{\xi^{1+\alpha}}
    \brr{
    	e^{-ik\xi}
        +e^{ik\xi}
        - 2
    }
    \widetilde{P}\sof{k;t}
    ,
\end{align}
or more compactly $\dot{\widetilde{P}}\sof{k;t} = S\sof{k}
    \widetilde{P}\sof{k;t}$, where
\begin{align}
		S\sof{k} 
		= 
		K
    \brr{
      \PolyLog{1+\alpha}{e^{-ik}}
      +
      \PolyLog{1+\alpha}{e^{ik}}
      -2 \PolyLog{1+\alpha}{1}
    }.    
    \nonumber
\end{align}
Here $\PolyLog{\nu}{z} = \sSum{n=1}{\infty} z^n/n^\nu$ is the polylogarithm function and $S\sof{k}$ is the Fourier symbol  of the transport operator defined as
\begin{align}
\widetilde{P}_x\sof{t} = \frac{1}{2\pi}\int dk \ e^{-ikx} e^{S\sof{k} t}.
\end{align}
Using the polylogarithm's expansion around $k=0$ (obtained by Mathematica) one finds the following small wave-vector expression for $S\sof{k}$:
\begin{equation}
S\of{k} 
\AsymEq 
2 \sGma{-\alpha} \cos\left(\frac{\pi \alpha}{2}\right) K \sAbs{k}^\alpha
.
\end{equation}
Note that the sign of $S\sof{k}$ is  negative for all $\alpha \in (0,2)$.
As we now have $\dot{\widetilde{P}}\sof{k;t} = S\sof{k} \widetilde{P}\sof{k;t}$, 
the solution is given by $\widetilde{P}\sof{k;t} = \exp\soff{S\sof{k}t} \AsymEq \exp\sof{-a \sAbs{k}^\alpha}$, 
where we used the initial condition $\widetilde{P}_x\sof{t=0} = \delta_{x,0}$, which gives $\widetilde{P}\sof{k;t=0} = 1$ and identified 
\begin{equation}
a = 2 \sGma{-\alpha} \cos\left(\frac{\pi \alpha}{2}\right) Kt .
\label{defa}
\end{equation}
The expansion also shows that $\widetilde{P}_x\of{t}$ is asymptotically equal to a symmetric stable distribution whose Fourier transform is exactly given by our stretched exponential.
The PDF of such a random variable decays like a power law for large $\sAbs{x}$, \cite{Feller1971-0}:
\begin{equation}
	\mathcal{F}^{-1}\offf{e^{-a\sAbs{k}^\alpha};x}
    \AsymEq
    \sGma{\alpha+1}
    \frac{\sin\sof{\alpha\pi/2}}{\pi} 
    \frac{a}{\sAbs{x}^{1+\alpha}}.
\end{equation}
Here $\mathcal{F}^{-1}$ denotes the inverse Fourier transform.
Using this equation and the definition of $a$ Eq.~\eqref{defa}, with $\Gma{\alpha+1} = \alpha \Gma{\alpha}$, $2\sin\of{y}\cos\of{y} = \sin\of{2y}$ and $\Gma{y}\Gma{-y} = \pi / \sin\of{\pi y}$, we recover Eq.~\eqref{eq:EMAAsym} from the main text. 

\bibliography{Article,Book,NotRead,Self}

\end{document}